\documentclass{mn2e}

\usepackage{psfig}

\def\ltsima{$\; \buildrel < \over \sim \;$}
\def\lsim{\lower.5ex\hbox{\ltsima}}
\def\gtsima{$\; \buildrel > \over \sim \;$}
\def\gsim{\lower.5ex\hbox{\gtsima}}

\begin{document}
\title[SGR flares and short GRBs]{SGR giant flares in the BATSE
short GRB catalogue: constraints from spectroscopy}

\author[Lazzati, Ghirlanda \& Ghisellini] 
{Davide Lazzati$^1$, Giancarlo Ghirlanda$^2$ and Gabriele Ghisellini$^2$ \\
$^1$ JILA, University of Colorado, Boulder, CO 80309-0440, USA {\tt e-mail:
lazzati@colorado.edu} \\ 
$^2$ Osservatorio Astronomico di Brera, via E. Bianchi 46, 
I-23807  Merate (LC), Italy \\ 
{\tt e-mail: ghirla;gabriele@merate.mi.astro.it} } 

\maketitle

\begin{abstract}
The giant flare observed on Dec. 27$^{\rm{th}}$ 2004 from SGR~1806--20
has revived the idea that a fraction of short ($<$2 s) Gamma Ray
Bursts (GRBs) is due to giant flares from Soft Gamma Ray Repeaters
located in nearby galaxies.  One of the distinguishing characteristics
of these events is the thermal (black body) spectrum with temperatures
ranging from $\sim$50 to $\sim$180 keV, with the highest temperature
observed for the initial 0.2~s spike of the Dec. 27$^{\rm{th}}$ 2004
event.  We analyzed the spectra of a complete sample of short GRBs
with peak fluxes greater than 4 photon s$^{-1}$ cm$^{-2}$
detected by BATSE.  Of the 115 short GRBs so selected only 
76 had sufficient signal to noise to allow the spectral analysis.  We
find only 3 short GRBs with a spectrum well fitted by a black body,
with $60\lsim{}kT\lsim90$~keV, albeit with a considerably longer
duration (i.e. $\gsim$1 sec) and a more complex light curve than the
Dec. 27$^{\rm{th}}$ 2004 event.  This implies a stringent limit on the
rate of extragalactic SGR giant flares with spectral properties
analogous to the Dec. 27$^{\rm{th}}$ flare.  We conclude that up to 4
per cent of the short GRBs could be associated to giant flares
($2\sigma$ confidence).  This implies that either the distance to
SGR~1806--20 is smaller than 15~kpc or the rate of Galactic giant
flares is lower than the estimated $0.033$~yr$^{-1}$.
\end{abstract}


\section{Introduction}

At 21:30 UT on December 27$^{\rm{th}}$ 2005 a giant flare of
$\gamma$--rays from the soft gamma repeater (SGR) SGR~1806--20 was
detected by INTEGRAL (Borkowski et al. 2004). The flare was detected
by all the active $\gamma$--ray observatories (Hurley et al. 2005;
Palmer et al.  2005) and by particle detectors (Schwartz et
al. 2005; Terasawa et al.  2005). SGRs are very likely hyper--magnetic
isolated neutron stars (Thompson \& Duncan 1995, 1996), with
$B\gsim10^{15}$~G. They are believed to be related to anomalous X--ray
pulsars (AXPs; Mereghetti \& Stella 1995), a class of neutron stars
with anomalously high X--ray luminosity and spin down rates.  AXPs and
SGRs share also an inferred magnetic field in the
$10^{14}\lsim{}B\lsim10^{15}$~G range.

\begin{figure*}
\centerline{\psfig{file=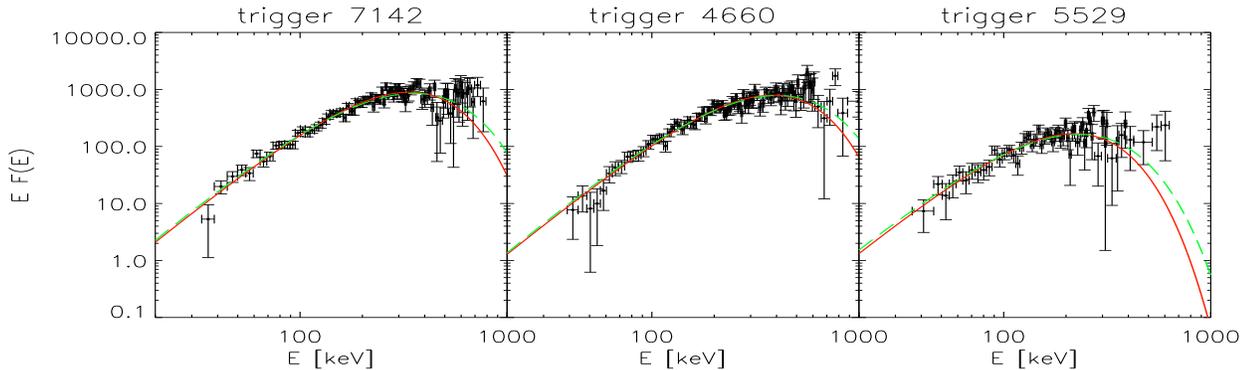,width=0.9\textwidth,height=4.8cm}}
\caption{Unfolded $\nu F_{\nu}$ spectra of the trigger 7142, 4660 and
5529. The red (green-dashed) line represents the best fit black body
(cutoff--powerlaw) model to the data points. Parameters of the best
fit spectral models are reported in Tab.~\ref{tab:1}.
\label{fig:spe}}
\end{figure*}

The flare followed a weak precursor (Hurley et al. 2005;
Mereghetti et al. 2005), with a spectrum similar to the one of
the recurrent weak bursts that characterize SGRs, but with much longer
duration. After the initial huge spike, which lasted $\sim200$~ms, a
roughly exponential tail of emission  characterized by the presence of a
deep periodic modulation at the spin period of the neutron star
$T=7.56$~s was observed for $\sim400$~s. A 
radio afterglow was detected on a week time
scale (Cameron et al. 2005; Gaensler et al. 2005), possibly
accompanied by X--ray emission (Mereghetti et al. 2005).

The Dec. 27$^{\rm{th}}$ flare is not the first giant flare detected
from an SGR. In 1979, a giant flare was detected from SGR~0525--66 in
the large Magellanic cloud (Mazets et al. 1979). Due to the saturation
of the instruments only a lower limit of few$\times10^{44}$~erg to the
energetics of the flare was derived. On August 27$^{\rm{th}}$ 1998 a
giant flare was detected from SGR~1900+14, for a total energetics of
$2\times10^{44}$~erg (Hurley et al. 1999). The recent flare from
SGR~1806--20 stands out in terms of its energetics. The initial spike
contains $\sim4\times10^{46}$~ergs, approximately a factor of 100 more
than in any previous recorded event. The peak flux was $2.5\times10^7$
photons~cm$^{-2}$~s$^{-1}$ (Terasawa et al. 2005) giving a luminosity
$L_{\rm{pk}}=2\times10^{47}$~erg~s$^{-1}$ (Hurley et al. 2005). Such
estimate relies on the
assumed distance of $D=15$~kpc to SGR~1806--20 (Corbel et al. 1997;
Corbel \& Eikenberry 2004; but see Figer et al. 2005, Cameron et
al. 2005; McClure-Griffiths \& Gaensler 2005).

Such a brightness makes extragalactic SGR giant flares (SGRGFs)
interesting as a possible source of short GRBs (Duncan 2001; Eichler
2002; Hurley et al. 2005; Nakar et al. 2005; Palmer et al. 2005).
Assuming a threshold count rate of $0.5$~photons~cm$^{-2}$~s$^{-1}$ in
the $[50-300]$~keV band for the BATSE trigger we derive, convolving a
black body spectrum with $kT=175$~keV to the BATSE response matrix, a
maximum distance of $D=35$~Mpc out to which an SGR with peak
luminosity $L=2\times10^{47}$~erg~s$^{-1}$ can be detected.  This
distance includes the Virgo Cluster of Galaxies (see Popov \& Stern
2005). Adopting the blue luminosity density normalization from Hurley
et al. (2005) and considering the BATSE sky coverage, we derive an
expected extragalactic SGRGF detection rate of:
\begin{equation}
\dot{N}_{\rm{SGRGF}}=30 \, (30/\tau)\,D_{15}^3 \qquad\qquad{\rm
yr}^{-1}
\label{eq:rate}
\end{equation}
where $\tau$ is the average time interval between the detection of two
Galactic SGRGF and $D_{15}$ is the distance to SGR~1806--20 in units
of 15~kpc.  This implies that approximatively 60 per cent of the
$\sim$ 500 BATSE short GRBs (defined as GRBs with $T_{90}<2$~s with
peak flux exceeding 0.5 phot cm$^{-2}$ sec$^{-1}$) should be
associated to SGRGFs.

It is possible to recognize SGRGF candidates in the BATSE catalogue
with different techniques. First, all the SGRs we know are inside
Galaxies and, since extragalactic SGRGFs can be detected up to a
relatively small distance, most of them should be associated to
bright star forming galaxies\footnote{Note that one of the
three known giant flares comes from the Large Magellanic
Cloud.}. Nakar et al. (2005, see also Hurley et al. 2002) studied 5
short GRBs error boxes, well localized by the interplanetary
network. They do not find any association, implying a loose upper
limit of less than $50\%$ ($2\sigma$) of SGRGFs in the BATSE short GRB
catalogue. Popov \& Stern (2005) apply an analogous technique by
looking for possible associations of short GRBs with high star
formation galaxies and the Virgo cluster. They do not find any
association. They conclude that the rate of
$E=10^{44}$~erg flares in our Galaxy should be less than one every 25
years and that the rate of giant $E=10^{46}$~erg flares should be less
than one in 1000 years.

In this letter we use the spectral diversity of SGRGFs with
respect to GRBs to search for candidate extragalactic SGRGFs in the
BATSE short GRB catalogue. The Dec. 27$^{\rm{th}}$ flare had a black
body spectrum with average $kT=175$~keV. Such a spectrum is harder
than a non-thermal GRB spectrum and can be singled out from
non--thermal GRB spectra even with a moderate number of counts.

\section{Spectral Analysis}

We have selected in the sample of 497 short duration GRBs (i.e.
$T_{90\%}<2$ sec - see also Magliocchetti et al.  2002) detected by
BATSE between May 2, 1991 and May 25, 2000 all the bursts with a peak
flux larger than 4 phot cm$^{-2}$ s$^{-1}$ in the energy range 50--300
keV (on the 64 ms timescale).  This cut was chosen in order to
assure enough signal to noise for a meaningful spectral analysis. More
importantly, it is the lowest flux for which spectral analysis of the
events can be performed irrespectively of their $T_{90}$ duration.  In
some cases the spectral analysis was not possible, but this was due
either to high background or non optimal detection configuration,
without any correlation to the burst duration.  We selected in this
way 115 GRBs: their durations are between 0.034 and 1.98 sec,
with an average of 0.6 sec, and their average fluence (integrated
above 25 keV) is of 2.8$\times 10^{-6}$ erg cm$^{-2}$.

We analyzed the BATSE (LAD) data of these bursts with the standard
methods (e.g. Preece et al. 2000).  Only for 76 GRBs we could extract
an average spectrum with enough signal to perform the spectral
analysis.  We fitted both the cutoff--powerlaw (CPL) model, that fits
well the spectrum of the 28 brightest short BATSE bursts (Ghirlanda,
Ghisellini \& Celotti 2004, which are included in the sample of
76 analyzed), and a black body (BB) spectrum representative of the
class of SGRs.

We found 3 out of 76 GRBs whose time integrated spectrum is well
fitted with a black body model and 15 events that does not exclude a
possible fit with a black body model, i.e.  yields a reduced
$\chi^{2}<1.5$.  This admittedly large $\chi^{2}$ value clearly
selects also those spectra with a very low signal to noise which does
not exclude the BB fit as well as any non--thermal model.  However, 12
of these events present systematic residuals at low and/or high
energies when fitted with a BB model.  Adding a non--thermal
component (i.e. powerlaw, PL) to their spectra improves significantly
the fit in most cases. The F--test probability of the null hypothesis
that the fit improves when adding model components ranges between
$10^{-2}$ ($2.5\sigma$) and $10^{-13}$ ($\gsim7\sigma$) for these 12
cases.  Nine of these 15 GRBs fitted with the BB+PL model present a BB
temperature not larger than 120 keV. Little more can be said for the
remaining 6 due to the low signal to noise.  Since the best fit
non--thermal component includes a sizable fraction of the flux, we
consider these spectra intrinsically different from the pure black
body spectrum of the giant flare from SGR~1806--20. However,
models with more than two degrees of freedom were not compared to the
data of SGR~1806-20, and so such mixed models ought to be considered
plausible.  Guidorzi et al.  (2004) find that the BeppoSAX data of
SGR~1900+14 cannot be fitted with a black-body spectrum.  In
addition the best--fit BB model temperatures are always much smaller
than the $kT=175$~keV observed in the giant flare.  Similar spectral
fits (BB+PL) have been performed and found acceptable for long GRBs as
well (Ghirlanda, Celotti \& Ghisellini 2003; Ryde 2004, Ghirlanda et
al.  2005).  In the next section we concentrate on the three remaining
cases, for which a fully consistent black body spectrum is found. We
also analyze in more detail three BB+PL bursts with interesting
characteristics.

\begin{figure}
\psfig{file=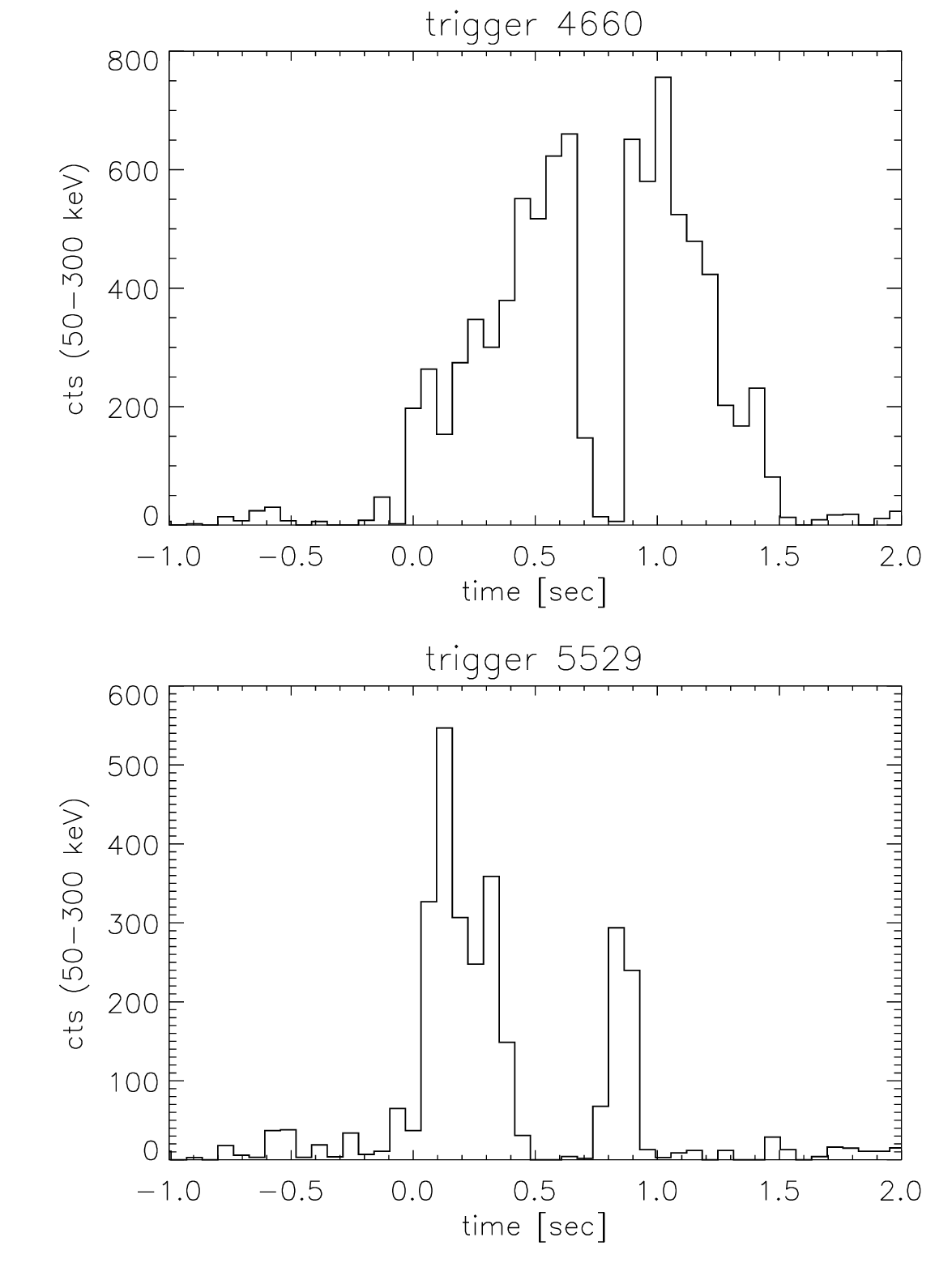,width=\columnwidth,height=6cm}
\caption{Background subtracted light curve in the energy range 50-300
keV for triggers 4660 and 5529, two of the candidates with thermal
spectrum. No 64~ms resolution data for trigger 7142 are available.
\label{fig:lc}}
\end{figure}

\begin{figure*}
\centerline{\psfig{file=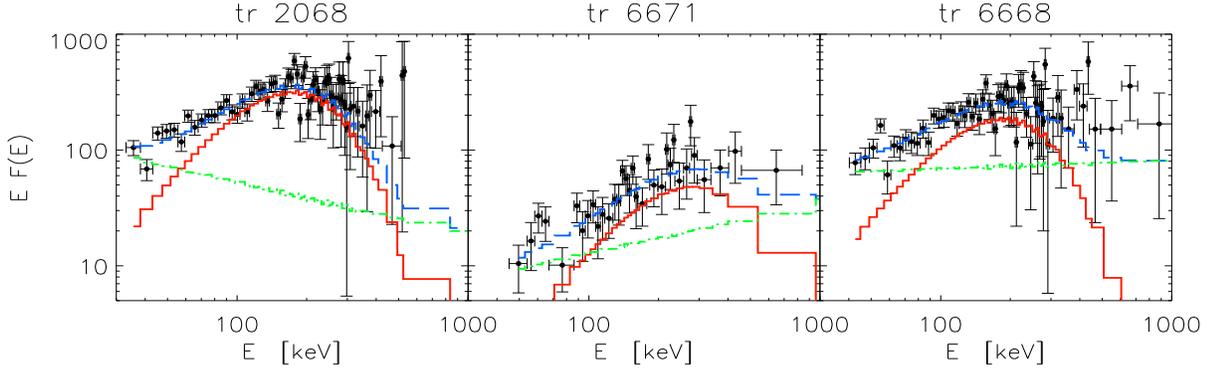,width=0.9\textwidth,height=4.8cm}}
\caption{$\nu F_{\nu}$ spectra of the trigger 2068, 6671 and 6668. The
dashed blue line is the best fit obtained with the combined models
BB+PL. Also shown are the two model components BB (red line) and PL
(greed dot--dashed line). Spectral parameters are reported in the
second part of Tab.~\ref{tab:1}.
\label{fig:spe2}}
\end{figure*}

\section{Candidate SGR giant flares}

In Tab.~\ref{tab:1} we report the basic properties of the 3 short GRBs
candidates showing evidence of a black body spectrum. As shown in the
table and in Fig.~\ref{fig:spe}, the fit with a non--thermal CPL model
results in a low energy photon spectral index $\alpha\sim$1.0 which
indicates an extremely hard spectrum when compared to the typical
values found in the class of bright long bursts (Ghirlanda, Celotti \&
Ghisellini 2002, 2003; Preece et al. 2000).

The spectra of trigger 7142 and 4660 are well fit by a black body
model with a characteristic temperature $kT$ of 86 and 98 keV,
respectively. The normalization of these two bursts is also similar.
A lower temperature ($kT=56$~keV) and normalization is instead found
for trigger 5529. We also verified the robustness of these fits
applying the same tests proposed in Ghirlanda et al.  (2003).  In
particular for the 3 bursts: (i) we selected different time intervals
to compute the GRB background, (ii) we analyzed the spectra of 2nd
brightest illuminated LAD detector and (iii) we used a different
Detector Response Matrix (of a GRB observed near the position of these
bursts but in a different epoch). Given the difficulty in the spectral
analysis to account for the sources of systematic errors, these tests
reinforce the result that the spectrum of these 3 bursts is indeed the
one presented in Tab.~\ref{tab:1} and Fig.~\ref{fig:spe}. This is the
first evidence of black body spectra also in the class of short GRBs
(see Ghirlanda et al. 2003, Ryde et al. 2004 for long GRBs).

The fact that these three events have a black body spectrum does not
necessarily imply that they are SGRGFs. It is indeed strange that the
only few candidates are found in the intermediate temperature range,
since, assuming a simple luminosity function of SGRGFs (e.g. a
power-law), either the low or high temperature ends should dominate
the statistics. Several consistency checks can be performed. We first
take a look at their light curves. These are shown in
Fig.~\ref{fig:lc} for the 4660 and 5529 triggers with 64~ms
resolution. No publicly available 64~ms resolution light curve for
trigger 7142 could be found. The two light curves are characterized by
two spikes of emission, well separated by a quiescent interval of
around 100~ms. A similar behaviour seems to characterize the 7142 low
resolution light curve posted on the BATSE website\footnote{\tt
ftp://cossc.gsfc.nasa.gov/compton/data/batse\\
/trigger/07001\_07200/07142\_burst}.  Such light curves are different
from the initial spikes of giant flares that have some variability but
at a moderate level (Terasawa et al. 2005). Also the duration of the
three candidates, of the order of one second, is longer than the
typical $\sim200$~ms observed in SGRGFs. No suitable thermal
spectrum candidate can be found for a GRB with duration around
$0.2$~s, where all the SGRGFs have been observed to date.

If we relax the hypothesis that the spectrum of short GRBs -- SGRGF
candidates should be a pure BB, we might still find candidates in the
15 GRBs whose $\chi^2$ does not exclude the BB model.  We further
limited the search among these 15 short bursts to those cases with a
single peaked light curve lasting less than $\sim$ 0.6 sec (i.e.  2.5
longer than SGR~1806-20).  We found 3 candidates, i.e.  trigger 2068,
6671 and 6668, whose spectral parameters are reported in the second
part of Tab.~\ref{tab:1}. In these cases the spectrum deviates from a
pure BB model at low and/or high energies and can be better
represented by a two component model, i.e.  adding a PL to the BB
(spectral parameters of the pure BB and BB+PL are reported in the
first and second line of each entry in Tab.~\ref{tab:1}).  The F-test
probabilities for the BB+PL model with respect to the BB model are
7$\times 10^{-5}$, $10^{-2}$ and 5$\times 10^{-5}$ for trigger 2068,
6671 and 6668, respectively.  The spectra of these 3 bursts and the
best BB+PL model are shown in Fig.~\ref{fig:spe2}.

\begin{table*}
\centering
\begin{tabular}{@{}lllllllllllll@{}}
\hline
Trigger & RA & Dec & $\delta (2\sigma)$ &  $t_{90}$ & $P$ &  $\alpha$  & 
$E_{\rm peak}$  & $\chi^2_r$ (dof) & kT  & $N$ & $\chi^2_r$ (dof)\\
        &    &     &          & s     & ph/(cm$^2$ s)  &            &
[keV]           &                        & [keV] &   &                     \\
\hline
7142 & 275.23 & $+$44.05 & 2.0 & 0.97$\pm$0.06 & 5.80$\pm$0.28 &  
1.07$_{-0.36}^{+0.14}$ & 355$\pm$61 & 0.95 (107)  & 86$_{-3.8}^{+4.0}$ 
&23.26$_{-1.37}^{+1.44}$ & 0.97 (108)\\
4660 & 305.83 & $-$40.27 & 1.9 & 1.17$\pm$0.08 & 5.15$\pm$0.30  &  
1.09$_{-0.45}^{+0.24}$ & 405$\pm$66 & 0.9 (101) & 98$_{-5.7}^{+6.3}$ & 
20.58$_{-1.65}^{+1.82}$ & 0.93 (102)\\
5529 & 29.45 & $-$0.29 & 4.5 & 1.02$\pm$0.13 & 4.23$\pm$0.29  &  
1.05$_{-0.26}^{+0.71}$ & 231$\pm$67 & 0.97 (101) & 56$_{-4.5}^{+6.0}$ & 
4.34$_{-0.45}^{+0.47}$ & 0.96 (102)\\  
\hline
2068 & 163.69 & $-$33.58 & 3.5 & 0.591$\pm$0.06 & 15.6$\pm$0.6 &
...  & ... & ... & 37$_{-1.9}^{+2.0}$ & 9.62$_{-0.51}^{+0.52}$ & 1.5 (108)\\ 
     & ...    & ...      & ...  & ...            & ...          &
-2.45$_{-0.34}^{+0.6}$  & ... & ... &  44$_{-3.5}^{+3.2}$ & ...  & 1.27 (106)\\
6671 & 191.36 & $-$44.93 & 8.2 & 0.256$\pm$0.091 & 6.7$\pm$0.3 &
...  & ... & ... & 62$_{-10}^{+11}$ & 1.87$_{-0.34}^{+0.41}$ & 0.92 (105)\\ 
     & ...    & ...      & ...  & ...            & ...          &
-1.6$_{-0.4}^{+3.6}$  & ... & ... &  71$_{-21}^{+34}$ & ...  & 0.88 (103)\\
6668 & 224.91 & $-$24.01 & 4.5 & 0.116$\pm$0.006 & 39.1$\pm$0.6 &
...  & ... & ... & 41$_{-3}^{+3}$ & 7.24$_{-0.54}^{+0.57}$ & 1.3 (108)\\ 
     & ...    & ...      & ...  & ...            & ...          &
-1.92$_{-0.3}^{+1.1}$  & ... & ... &  47$_{-8}^{+8}$ & ...  & 1.1 (106)\\
\hline \hline
\end{tabular}
\caption{Spectral parameters of the 3 short GRBs with average spectrum
found fully consistent with a black body model.  $\delta$: positional
error in deg; $P$: 64 ms peak flux in the energy range 50--300 keV;
$\alpha$: Low energy photon spectral index of the cutoff--powerlaw
spectral model; $E_{\rm peak}$: peak energy of the $\nu F_{\nu}$
spectrum of the cutoff--powerlaw model; $N$ Normalization of the black
body model in units of $L_{39}/D_{10}^2$ where $L$ is the burst
luminosity in units of 10$^{39}$ erg s$^{-1}$ and $D_{10}$ is the
burst distance in units of 10 kpc.  Also listed are the 3 short GRBs
for which the BB fit is not excluded, with duration $<$0.6 s and a
single peaked light curve. In the last three cases the fit improves
when using the BB+PL model. The second line of trigger 2068, 6671,
6668 represents the fit parameters of the BB+PL model (F--test
probabilities are given in the text).
\label{tab:1}}
\end{table*}

\subsection{Counterparts}

If the emitting surface of the 3 short GRB candidates is similar to
that of the SGR~1806--20 and their luminosity scales as
$L\propto(kT)^{4}$, as expected in a baryon free fireball model, it is
possible to infer their luminosities: $L_{7142}=1.1\times
10^{46}$~erg/sec, $L_{4660}=1.74\times 10^{46}$~erg/sec and
$L_{5529}=1.9\times 10^{45}$~erg/sec.  These luminosities allow us to
compute the putative distance to these GRBs, given the 50--300 keV
flux as derived from the best fit of the spectra. This can be slightly
different than the peak flux reported in the BATSE catalogue which is
calculated with a low resolution spectrum (4 energy channel data) via
direct inversion techniques (e.g.  Fishman et al.  1994). We find that
the distance should be $d_{\rm GRB}\propto d_{\rm lim,GRB} (P_{\rm
lim}/P)^{1/2}$, where $P$ is the 50--300 keV integrated model flux.
We find distances of 3.14 Mpc, 4.9 Mpc and 2.33 Mpc for trigger 7142,
4660 and 5529, respectively.

We have searched the NED database for galaxies within the $2~\sigma$
error box\footnote{The $2\sigma$ error box was derived from the
BATSE catalog error box assuming a circular shape and Gaussian
statistics.} of the 3 short GRB candidates. 
We also allowed for a larger error box with
respect to the $\delta$ values reported in Tab.~\ref{tab:1} by adding
the systematic BATSE positional uncertainty of $\sim 1.6^\circ$. In
all cases but trigger 4660 we find more than one galaxy with
$z\le0.0082$ within the error box. In no case, however, we find a
galaxy within the distances derived above. Where a positive
result would have been interesting, this negative result does not
allow us to derive any constraint given the complex shape and
statistics of the BATSE error areas and the fact that the SGRGF may
originate from small low-surface galaxies, such as the Large
Magellanic Cloud. Analogous conclusions can be derived for the three
selected bursts with a BB+PL spectrum.

\section{Discussion}

We have analyzed the spectra of the 76 among the 115 brightest BATSE
short GRBs, defined as GRBs with $T_{90}<2$~s and 50--300 keV peak
flux $P \ge 4$ phot cm$^{-2}$ s$^{-1}$.  We searched for black body
spectra (possibly with high temperature $kT\gsim150$~keV) to identify
candidate extragalactic SGRGFs.  We find 3 candidate black body
spectra, albeit with lower temperature.  If we assume that the
luminosity scales as $L\propto(kT)^4$, such events should be at a
distance of $d\sim$2--5~Mpc.  Their multi--peaked light curves and
durations longer than 1 second argue against their identification as
SGRGFs.  We therefore conclude that we could not find any robust
evidence of the presence of SGRGFs in our sample of short BATSE GRBs.
The expected number of SGRGFs can be computed from
Eq.~\ref{eq:rate} rescaling the threshold from $0.5$ to $4$
photons~cm$^{-2}$~s$^{-1}$. For the 9.5 years of BATSE activity we
find:
\begin{equation}
N_{\rm{SGRGF}} = 12.6 \,(30/\tau) D_{15}^3
\end{equation}
The probability of finding no event out of an expected number of 12.6
is given by Poisson statistics and corresponds to $P=3.4\times10^{-6}$
or $\sim4.5\sigma$. Alternatively we can give an upper limit
$N_{\rm{SGRGF}}<3$ at the $2\sigma$ level.  Several factors can
contribute to this mismatch.

The extragalactic SGRGF rate depends on the distance to
SGR~1806--20 to the third power (Eq.~\ref{eq:rate}; Hurley et
al. 2005; Nakar et al. 2005). A distance of
$\lsim$9~kpc would be consistent with both our non detection and with the 
radio absorption measurements (Cameron
et al. 2005; McClure--Griffiths \& Gaensler 2005). It would place
the SGR outside the massive star cluster with which is associated in
projection. Such smaller distance would decrease the SGR~1806-20 flare
luminosity by a factor $\sim5$.
BATSE trigger criteria underwent many changes during the BATSE
activity and we have adopted an average effective value.  This,
however seem not enough to explain such a big discrepancy (our
estimated rate is conservative).
The rate of Galactic SGRGFs may be smaller than estimated. 
If $\tau\gsim130$~yr, the failed detection of
extragalactic flares would be marginally acceptable. We should then
conclude that we happened by chance to live in a period when a
Galactic SGRGF went off.
A final possibility is that the SGRGF spectrum is not purely
black body. We find several spectra that can be modelled with a black
body component plus a power--law. Three of them  satisfy
also constraints on the light curve duration and variability. If a
sizable non thermal emission contributes to the SGRGF emission at low
and high frequencies, these short GRBs may be good candidates for an
association.
The discrepacy would not however be solved since we find only 3 events
out of the expected 13.

The solid conclusion of our analysis, similar to what independently
derived by Popov and Stern (2005) with positional association
techniques, is that the rate of $L>10^{47}$~erg~s$^{-1}$ giant flares
is smaller than one every $\sim130$ years\footnote{Their constrain
is somewhat tighter, but note that due to the complex shape and non
Gaussian statistics of BATSE error boxes, their analysis may be
heavily affected by systematic uncertainties.}.  It is not clear
whether there is any such luminous giant flare in the BATSE sample.
If we ask instead what is the rate of giant flares similar to the
SGR~1806--20 one, we face the uncertainty on its luminosity, a consequence
of our ignorance of its exact distance. For a distance of 9~kpc, 
the peak luminosity of SGR~1806--20 would be
$L=10^{47}$~erg~s$^{-1}$, and the rate of BATSE detectable flares be
as low as 3 in the whole BATSE activity for our countrate
threshold, marginally consistent with our result.  

It is finally worth discussing some effects that could add small
uncertainty to our estimates. We checked that, with the adopted
threshold, there is no bias against short events in the spectral
analysis. Some residuals non statistically significant effects may
still be present but should be marginal.  On the other hand, the
effects of scattering of burst photons in the detector and atmosphere
are included in the response matrices (Preece et al. 2000), that
are computed case by case. The spectral shape should be therefore
unaffected by instrumental effects. We finally note that the spectra
are mostly constrained in the intermediate energy range, so that any
calibration uncertainty at the edge of the detector sensitivity will
not affect our results.

\section*{Acknowledgements}

We would like to thank our referee, Kevin Hurley, for the very
constructive comments and suggestions. We would like to thank Robert
Duncan and Ehud Nakar for useful discussions and comments. This
research has made use of the data obtained through the High Energy
Astrophysics Science Archive Research Center Online Service, provided
by the NASA/Goddard Space Flight Center. This work was supported in
part by NSF grant AST-0307502 and NASA Astrophysical Theory Grant
NAG5-12035 (DL). GG \& GG thanks the Italian MIUR for COFIN-Grant.

\end{document}